\documentclass[twocolumn,showpacs,prl,amsmath,amssymb]{revtex4}

\usepackage{graphicx}

\def\bra#1{\langle#1 |}
\def\ket#1{| #1\rangle}

\def\ops#1{\hat{#1}}

\def\diagonalav#1{\left<#1\right>_{\mathrm{D}} }

\newcommand{\ud}{ \mathrm{d} }

\begin{document}

\title{Relaxation Dynamics of Disordered Spin Chains: \\ Localization and the Existence of a 
Stationary State}

\author{Simone Ziraldo$^{1,2}$, Alessandro Silva$^{3}$, and Giuseppe E. Santoro$^{1,2,3}$} 
\affiliation{
$^1$ SISSA, Via Bonomea 265, I-34136 Trieste, Italy \\
$^2$ CNR-IOM Democritos National Simulation Center, Via Bonomea 265, I-34136 Trieste, Italy \\
$^3$ International Centre for Theoretical Physics (ICTP), P.O.Box 586, I-34014 Trieste, Italy
}

\begin{abstract}
We study the unitary relaxation dynamics of disordered spin chains following a sudden
quench of the Hamiltonian.
We give analytical arguments, corroborated by specific numerical examples, to show that the 
existence of
a stationary state depends crucially on the spectral and localization properties of the final 
Hamiltonian, and not on the initial state.
We test these ideas on integrable one-dimensional models of the Ising or $XY$ class, but argue 
more generally on their validity for more complex (nonintegrable) models.
\end{abstract}


\date{\today}
\maketitle

Ergodicity is a fundamental concept of classical mechanics: the properties of
dynamical trajectories in phase space determine the long-time dynamics of a system and the
description of the eventual stationary state in terms of statistical mechanics. 
The extension of these ideas to the quantum realm, pioneered in 1929 by 
J. von Neumann~\cite{vonNeumannHTheorem,vonNeumannTranslation}, has motivated a 
great deal of recent research, mainly spurred by the experimental possibility of
studying the nonequilibrium 
dynamics of thermally isolated quantum systems  -- most notably cold atomic 
gases in optical lattices~\cite{Bloch_RMP08,Lewenstein_AP07}. 
A highly debated issue in the recent literature is the characterization 
of the long-time dynamics of a quantum system taken out of equilibrium by a sudden change of 
one of its parameters (a quantum quench).
If an extensive amount of energy is suddenly injected in the system, will the resulting dynamics tend always to 
a well defined stationary state? And what is the statistical ensemble describing it?    

The stationary state existence, has been investigated 
both in generic systems~\cite{vonNeumannHTheorem} 
and in Hubbard-type models~\cite{Cramer2008,Cramer2010,Flesch2008}.
A fast dynamical relaxation was recently observed experimentally~\cite{Trotzky2012}
in a system of cold atoms and its long-time stationary state results were compatible with the
generalized Gibbs ensemble (GGE)~\cite{Jaynes_PR57,Rigol_PRB06},
where a set of macroscopic constants of motion are constrained by the initial state. 
Once a stationary state is established, the integrability or nonintegrability of the dynamics appears to be 
the crucial ingredient: 
while integrable systems in the thermodynamic limit
are often described by a GGE~\cite{Barthel2010,Calabrese2011,Cazalilla2012}, it is generally
expected that the breaking of integrability will lead to thermalization~\cite{Rigol_Nat}.

While the relaxation dynamics of uniform systems is well understood, recent
studies hinted towards nontrivial effects due to the breaking of translational invariance. 
In the debate two features emerged: the importance of distinguishing between 
thermodynamic limit~\cite{Calabrese2011,Cazalilla2012}  and finite-size effects~\cite{Gangardt2008,Caneva_JSM11}, 
and the possible role played by localization~\cite{Khatami}.  
For example, while breaking translational invariance in the initial state could introduce correlations 
among different constants of motion,
relevant for finite-size systems~\cite{Gangardt2008,Caneva_JSM11},
their effect has been argued to be negligible in predicting the stationary state attained by local 
observables~\cite{Cazalilla2012}. 
While in the thermodynamic limit the breaking of translational invariance may not have a significant effect, 
localization could in turn play an important role, to the extent of resulting in the absence of 
thermalization even in nonintegrable spin chains~\cite{Khatami}. This observation appears 
to be consistent with earlier numerical analysis in
disordered Ising or $XY$ spin chains (characterized by 
localization of the eigenstates), where a discrepancy between the expected GGE and the 
effective stationary state was  observed~\cite{Caneva_JSM11}. 

The purpose of this work is to characterize the long-time dynamics of 
disordered systems focusing 
on the existence of a stationary state for local observables. 
We do so by studying 
the time fluctuations of local observables in disordered spin chains which can be mapped to 
free fermionic systems.
On the basis of {\it analytical} calculations, corroborated by numerics, we argue that the existence of a stationary state depends crucially 
on the spectral properties of the {\em final} Hamiltonian, and not on the initial state. While for 
final Hamiltonians with a continuous spectrum associated to delocalized states 
a well defined stationary state is attained (in the thermodynamic limit), 
in systems with localized states (possessing a pure-point local spectrum) time-fluctuations of local 
observables persist.
This will be generally associated to a failure of the GGE in describing long-time averages of many-body
operators. 
While we have tested these ideas on integrable Ising or $XY$ models, we will argue 
for their validity for more complex (nonintegrable) systems.

%
\begin{figure}[tbh]
   \begin{center}
	\includegraphics[width=0.5\textwidth]{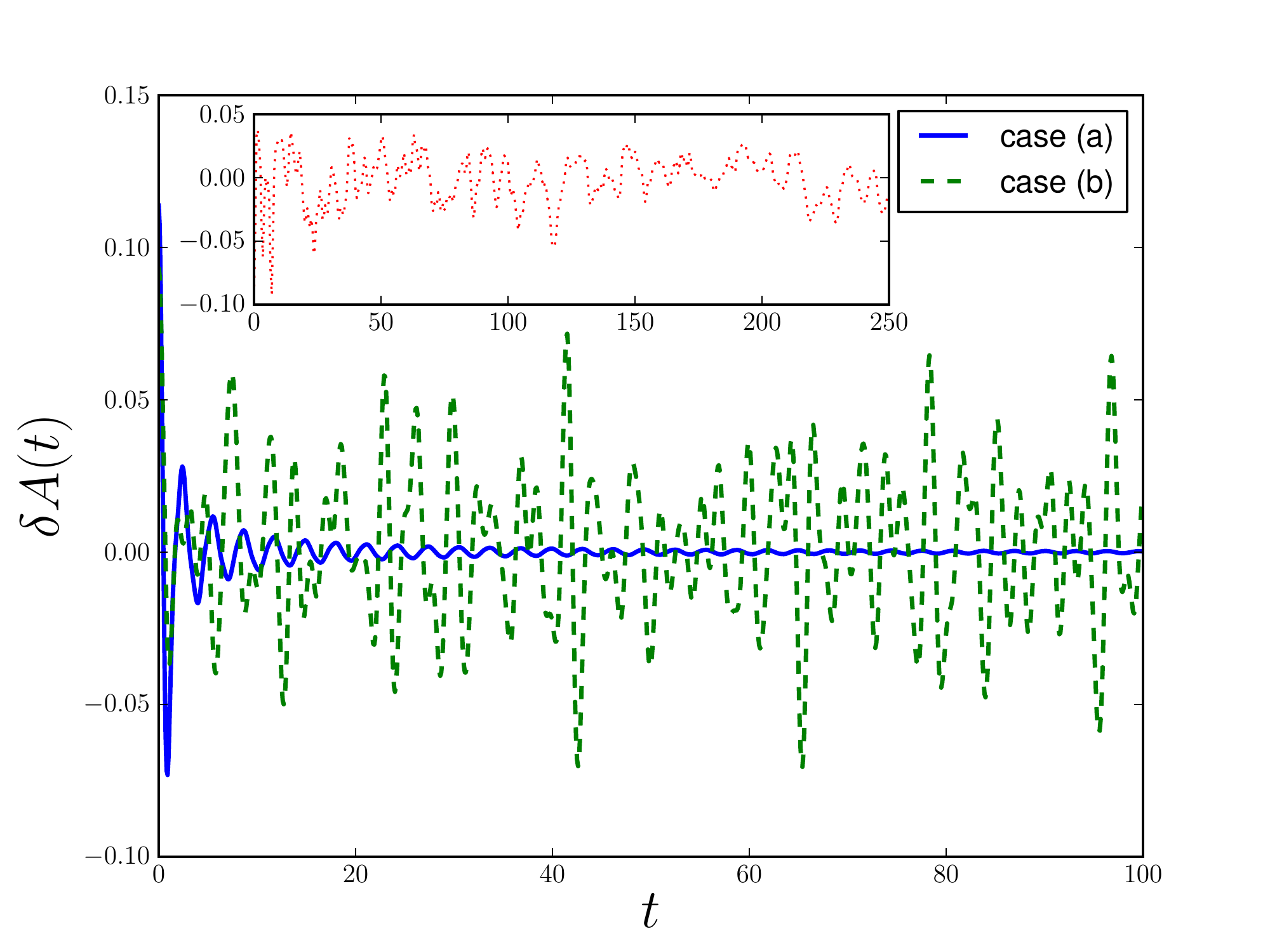}
   \end{center}
\caption{ \label{fig:te}
Illustration of the two possible behaviors for the time-dependence of an operator $A(t)$, 
showing the fluctuations around the time average, which decrease [case (a)] or persist at all 
times [case (b)]. Case (a) is a quench between disorder-free Ising chains ($h_0=0.9$, $h_f=0.5$), 
and case (b) a quench
from a disorder-free to a disordered $XX$ chain (with $\epsilon=0.3$, see text for details).
The inset shows an ambiguous case: a quench from a disordered to a clean $XX$ chain.
In all cases, $\hat{A} = \ops{c}_j^\dagger \ops{c}_j$, the local fermionic density operator.
%
}
\end{figure}
%

Let us start by defining the problem. Consider a standard quantum quench:  an initial state 
$|\psi_0\rangle$, ground state of some Hamiltonian $\ops{H}_0$, evolves under a different 
time-independent Hamiltonian $\ops{H}$.
Given an observable $\hat{A}$, its average can be separated in two terms: 
\begin{equation} \label{defA:eqn}
A(t) \equiv \langle \psi_0 | e^{i\ops{H}t} \hat{A} e^{-i\ops{H}t} | \psi_0\rangle = \bar{A} + \delta A(t)  \, ,
\end{equation}
where $\bar A \equiv \lim_{T\to \infty} \frac{1}{T} \int_0^T \! dt \; A(t)$ is its time average and 
$\delta A(t)$ is the fluctuating part.
Denoting by $|\alpha\rangle$ the eigenstates of $\ops{H}$ with energy $E_{\alpha}$, 
and defining $C_{\alpha}\equiv \langle \alpha | \psi_0\rangle$ and $A_{\alpha'\alpha} \equiv \bra{\alpha'} \hat{A} \ket{\alpha}$, we get
\begin{eqnarray}
A(t) &=& \sum_{\alpha} 
       |C_{\alpha}|^2 A_{\alpha\alpha} + 
        \sum_{\alpha'\ne \alpha} e^{i(E_{\alpha'}-E_{\alpha})t}
       C_{\alpha'}^* A_{\alpha'\alpha} C_{\alpha}  \nonumber \\
     &=& \diagonalav{\hat{A}} +   \int_{-\infty}^{+\infty} \! \! \! d\Omega \, e^{-i\Omega t} F_A(\Omega) \, , \nonumber
\end{eqnarray}
where $\diagonalav{\hat{A}}$ is the so-called 
{\em diagonal average} \cite{vonNeumannHTheorem,vonNeumannTranslation,Rigol_Nat}
which coincides with $\overline{A}$ (assuming no energy degeneracies), 
and the fluctuating part $\delta A(t)$ has been recast as
the Fourier transform of a weighted joint density of states 
$F_A(\Omega) \equiv \sum_{\alpha'\ne\alpha} C_{\alpha'}^* A_{\alpha'\alpha} C_{\alpha} \delta\left(\Omega-E_{\alpha}+E_{\alpha'}\right)$.
The behavior of the fluctuating part $\delta A(t)$  (see Fig.~\ref{fig:te}) decaying to 0 for large $t$, 
case (a), or remaining finite (with persistent oscillations), case (b), is strongly tied to the smoothness of $F_A(\Omega)$. 
If the many-body spectrum $\{E_{\alpha}\}$, in the thermodynamic limit, 
is a smooth continuum and the weights $C_{\alpha'}^* A_{\alpha'\alpha} C_{\alpha}$ make 
$F_A(\Omega)$ still integrable, then $\delta A(t)$ will decay to zero for large $t$, 
due to the destructive interference induced by the strongly oscillating phase $e^{-i\Omega t}$ (Riemann-Lebesgue lemma). 
If, on the contrary, $\{ E_{\alpha} \}$ has an important {\em pure-point spectrum} part, 
i.e., delta functions associated to localized eigenstates, 
then one should expect persistent time fluctuations for certain operators. 

To exemplify the previous general arguments, we consider specific quenches for  
Hamiltonians of the Ising or $XY$ type in one dimension:
\begin{equation}\label{eq:isingham}
  \ops{H} = - \sum_{j=1}^L \left( J_j^x \ops{\sigma}^x_j \ops{\sigma}^x_{j+1} 
                                                    + J_j^y \ops{\sigma}^y_j \ops{\sigma}^y_{j+1} \right) 
      - \sum_{j=1}^L h_j \ops{\sigma}_j^z \;,
\end{equation}
where $L$ is the size of the chain, $\ops{\sigma}_j^\mu$ ($\mu = x,y,z$) are Pauli matrices at
site $j$ with $\ops{\sigma}_{L+1}^\mu \equiv \ops{\sigma}_{1}^\mu$.
$J_j^x$, $J_j^y$ and $h_j$ are nearest-neighbor spin couplings and transverse magnetic fields.
By a Jordan-Wigner transformation \cite{Lieb_AP61}
$\ops{H}$ can be rewritten as a quadratic form of fermionic operators $\ops{c}_j$,
and through a Bogoliubov rotation we define a 
new set of fermionic operators which diagonalize $\ops{H}$ \cite{Lieb_AP61,Pfeuty}.
Wick's theorem allows us to express $A(t)$ in terms of sums of products of one-body Green's functions
${G}_{j_1j_2}(t) \equiv \langle \psi(t) | \ops{c}^\dagger_{j_1}\ops{c}_{j_2} | \psi(t) \rangle$ 
and ${F}_{j_1j_2}(t) \equiv \langle \psi(t) | \ops{c}^\dagger_{j_1}\ops{c}^\dagger_{j_2} | \psi(t) \rangle$.
Therefore, establishing that ${G}_{j_1j_2}(t)$ and ${F}_{j_1j_2}(t)$ approach a well-defined limit for 
large $t$ (i.e., their fluctuations decay) allows us to make similar
statements for a large class of operators, including spin-spin correlations 
$\ops{\sigma}^x_{j_1} \ops{\sigma}^x_{j_2}$, 
$\ops{\sigma}^z_{j_1} \ops{\sigma}^z_{j_2}$, and many others
which may be local or nonlocal in terms of the $\ops{c}_i$ fermions.
%
%
The case of quenches in a homogeneous Ising chain
($J_j^x = 1$, $J_j^y = 0$), with $h_j = h_0$ in $\ops{H}_0$, 
and $h_j = h_f \neq h_0$ in $\ops{H}$, is simple to illustrate: 
for $L\to \infty$, we have
$
\delta G_{j_1j_2}(t) = C \int_0^{\pi} \! \frac{dk}{2\pi} 
\frac{\cos \left[ (j_1-j_2)k \right] \sin^2 k}{\epsilon_{k,{\rm f}}^2 \epsilon_{k,{\rm 0}}} 
\cos \left( 2 \epsilon_{k,{\rm f}} t \right) 
$,
where $C \equiv 4(h_0-h_f)$ and 
$\epsilon_{k,{\rm 0/f}}\equiv\sqrt{1+2h_{\rm 0/f} \cos k + h^2_{\rm 0/f}}$ are
the quasiparticle energies for $\ops{H}_0$ and $\ops{H}$.
Case (a) of Fig.~\ref{fig:te} is an illustration for
$j_1=j_2$ (the local density), where fluctuations decay to 0 as $t^{-{3/2}}$ with 
oscillations~\cite{suzukiArxiv}.
This follows from the continuous single-particle spectrum and from
the Riemann-Lebesgue lemma.
A similar statement holds for $j_1 \neq j_2$ and for $F_{j_1j_2}(t)$. 

Now we turn to disordered Hamiltonians. To simplify the presentation 
we focus on quenches for transverse-field $XX$ chains, $J^x_j=J^y_j$, where BCS terms are absent. 
When disorder is present, 
simple analytical expressions are lacking, while pure numerics leads to ambiguous 
results.
Figure~\ref{fig:te}, for instance, shows results for $\delta G_{jj}(t)$ upon quenching  
from a clean $\ops{H}_0$ (with $J^x_j=J^y_j=1$ and $h_j=0$) to a disordered $\ops{H}$ 
with $J_j^x=J_j^y = 1 + \epsilon \eta_j$, $h_j = \epsilon \xi_j$ and $\epsilon = 0.3$
($\epsilon$ sets the disorder strength and $\eta_j$,  $\xi_j$ are 
uncorrelated uniform random numbers in $[-1, 1[$).
In this case, the fluctuations of $G_{jj}(t)$ are quite clearly persisting at all times.
The inset of Fig.~\ref{fig:te}, however, shows results obtained from the opposite 
quench (from a disordered $\ops{H}_0$ to a clean $\ops{H}$), and the 
result for $\delta G_{jj}(t)$ is now much more ambiguous 
(we prove below that fluctuations decrease to 0 in such a case).
To make progress we need a quantitative discrimination of the persistence of fluctuations and
we introduce the mean squared fluctuations of $G_{j_1j_2}(t)$:
\begin{equation} \label{eq:defDelta2}
\delta_{j_1j_2}^2 \equiv \lim_{T \rightarrow \infty}\frac{1}{T} \int_0^T \ud t \left| \delta G_{j_1j_2}(t) \right|^2 \; ,
\end{equation}
which is $0$ when time fluctuations of $G_{j_1j_2}(t)$ vanish [case (a)], and
finite if they persist [case (b)]. 
Physically, $\delta_{jj}^2$ is the 
average fluctuation of the local density $\ops{c}_j^\dagger \ops{c}_j$.
Let us denote by
$\ops{c}^\dagger_\mu  = \sum_{j=1}^L u_{j\mu} \ops{c}^\dagger_j$ 
the fermionic operators diagonalizing $\ops{H}$ and with $\epsilon_{\mu}$ their energy.
If $\ops{H}$ is translationally invariant then $\mu$ is the momentum $k$ and $u_{jk}=e^{ikj}/\sqrt{L}$, 
while if $\ops{H}$ is disordered $u_{j\mu}$ are localized eigenfunctions.
By expanding the $\ops{c}_{j}$'s in terms of the $\ops{c}_{\mu}$'s, we get
$G_{j_1j_2}(t) = \sum_{\mu_1\mu_2} e^{i (\epsilon_{\mu_1}-\epsilon_{\mu_2})t} u^\ast_{j_1 \mu_1} u_{j_2\mu_2}  G_{\mu_1\mu_2}$, 
where $G_{\mu_1\mu_2}=\bra{\psi_0} \ops{c}^{\dagger}_{\mu_1} \ops{c}_{\mu_2} \ket{\psi_0}$.
Assuming no energy degeneracy (i.e., $\epsilon_{\mu_1} = \epsilon_{\mu_2}$ only if $\mu_1 = \mu_2$)
$\delta G_{j_1j_2}(t)$ has the same expression as $G_{j_1j_2}(t)$, except for the
absence of the terms with $\mu_1=\mu_2$.  
The integrand in Eq.~\eqref{eq:defDelta2} is therefore
\begin{equation}\label{eq:doublesum}
\begin{split}
\left| \delta G_{j_1j_2}(t) \right|^2
= \sum_{\mu_1 \neq \mu_2} \sum_{\mu_3 \neq \mu_4}
e^{i (\epsilon_{\mu_1} - \epsilon_{\mu_2}  - \epsilon_{\mu_3} + \epsilon_{\mu_4} ) t} 
\\
u^\ast_{j_1 \mu_1} u_{j_2\mu_2} u_{j_1 \mu_3} u_{j_2\mu_4}^\ast
G_{\mu_1\mu_2} G_{\mu_3\mu_4}^\ast \;. \nonumber
\end{split}
\end{equation}
With the further assumption of no gap degeneracy  (i.e., $\epsilon_{\mu_1} - \epsilon_{\mu_2} = 
\epsilon_{\mu_3} - \epsilon_{\mu_4}$ only if  
$\mu_1 = \mu_3$ and $\mu_2 = \mu_4$, or $\mu_1 = \mu_2$ and $\mu_3 = \mu_4$) 
\cite{Reimann_note} we arrive at the key result
\begin{equation}  \label{eq:delta2}
\delta_{j_1j_2}^2 = \sum_{\mu_1 \neq \mu_2} |u_{j_1\mu_1}|^2 |u_{j_2\mu_2}|^2 
|G_{\mu_1\mu_2}|^2 \;,
\end{equation}
expressing $\delta_{j_1j_2}^2$ for a single realization as an 
eigenfunction-weighted sum of $|G_{\mu_1\mu_2}|^2$.
%
%
Disorder averages are performed {\em after} computing $\delta^2_{j_1j_2}$,
because we want to analyze the fluctuations of a given realization 
(averaging $\delta G_{j_1j_2}(t)$ would cancel such fluctuations).
Notice that, while eigenfunction properties are buried in the $C_{\alpha}$
and $A_{\alpha\alpha'}$ factors appearing in $F_{A}({\Omega})$,
a many-body theory of fluctuations starting directly from $F_{A}({\Omega})$ is 
difficult \cite{Reimann_note}.

The nature of the eigenfunctions (localized versus extended) plays a crucial role
in Eq.~\eqref{eq:delta2}.
Regardless of disorder, the $|G_{\mu_1\mu_2}|^2$'s sum to the total 
number of fermions $N_{\mathrm{F}}^0$ in the initial state:
\begin{equation}\label{G-average:eqn}
\sum_{\mu_1 \mu_2} | \bra{ \psi_0 } \ops{c}^{\dagger}_{\mu_1} \ops{c}_{\mu_2} \ket{ \psi_0 } |^2 = N_{\mathrm{F}}^0 \;.
\end{equation}
Therefore, if the final eigenstates are extended, $|u_{j_1\mu_1}|^2 |u_{j_2\mu_2}|^2 \sim1/L^2$, then
$\delta_{j_1j_2}^2$ in (\ref{eq:delta2}) scales to zero as  $N_{\mathrm{F}}^0/L^2\sim 1/L$ for a system with a finite density of fermions. 
If $\ops{H}$ is clean we have to take care of the degeneracy
$\epsilon_k=\epsilon_{-k}$ and the particle-hole symmetry, but still, see supplementary material (SM),
we can prove a bound $\delta^2_{j_1j_2} \leq  8N_{\mathrm{F}}^0/L^2$,  indicating that
fluctuations vanish for $L\to \infty$.
This generalizes the result of Ref.~\cite{Cazalilla2012} to cases where
$\ops{H}_0$ might lead to a $|G_{k_1k_2}|^2$ which has important
nonvanishing contributions also for $k_1\ne k_2$ (see SM).
%

\begin{figure}[tbh]
   \begin{center}
	\includegraphics[width=0.5\textwidth]{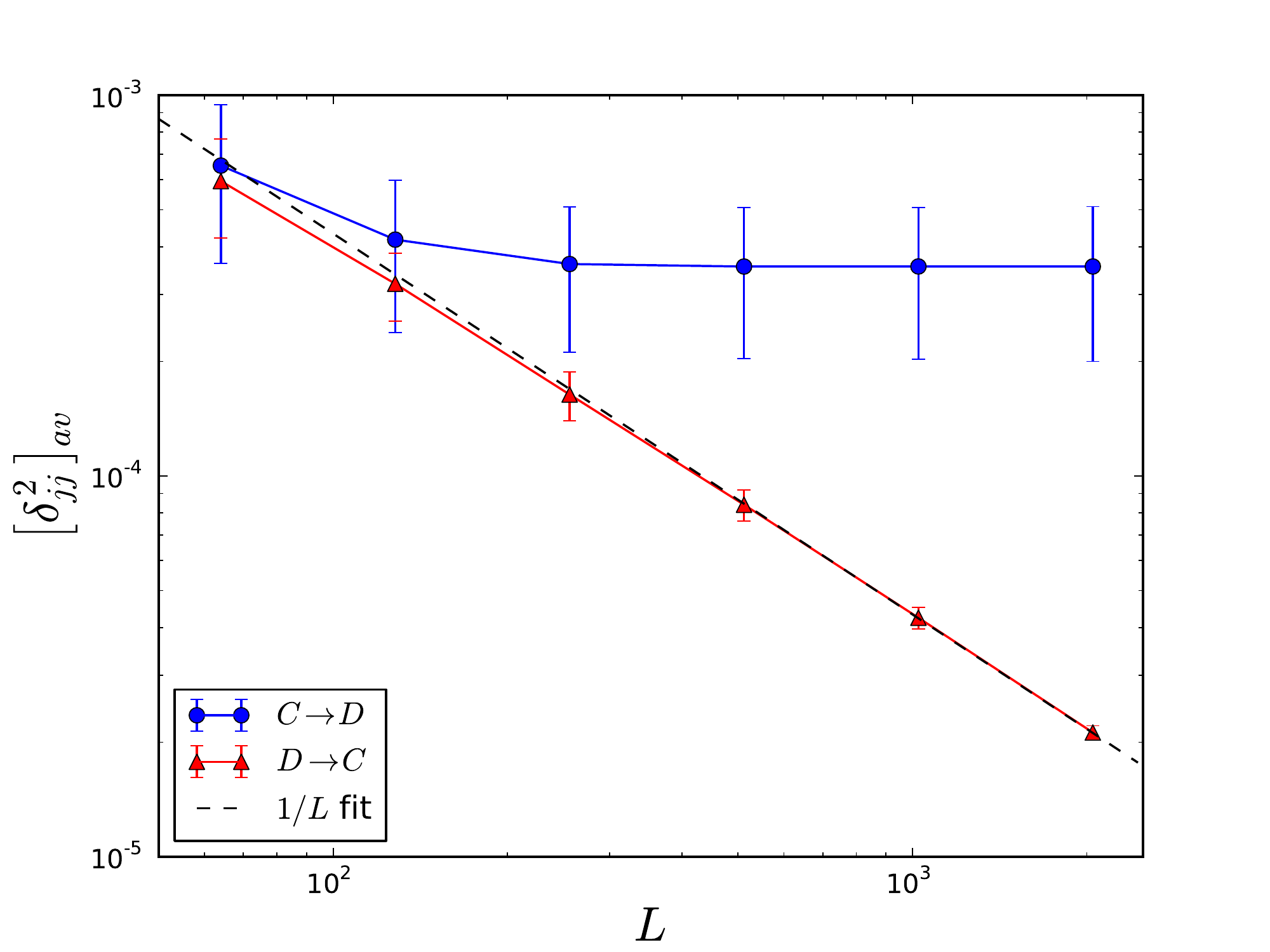}
   \end{center}
\caption{Average values of $\delta^2_{jj}$ with $j=L/2$, Eqs.~\eqref{eq:defDelta2} and \eqref{eq:delta2}, 
for quenches of $XX$ chains when $\ket{\psi_0}$ is the ground state 
of a disordered $\ops{H}_0$ (with $\epsilon=0.3$), while $\ops{H}$ is clean with $h_j=0$ (triangles $D \rightarrow C$), 
or  the opposite case (circles $C \rightarrow D$), see text for details.
The averages are taken over $200$ different disordered realizations 
and the error bar is the standard deviation of the distribution (not the error on the average). 
}
\label{fig:delta2}
\end{figure}

Figure~\ref{fig:delta2} shows the disorder average $[\delta^2_{jj}]_{av}$ 
as a function of $L$  in
the two opposite situations described above, i.e., quenches from 
a disordered $\ops{H}_0$ to a clean $\ops{H}$ ($D \to C$), or vice versa ($C \to D$).
In all cases $\delta^2_{jj}$ is calculated 
from Eq.~\eqref{eq:delta2} (with the extra terms due to degeneracies in the $D\to C$ case).
When $\ops{H}$ is clean ($D\to C$ data), $[\delta^2_{jj}]_{av}$ scales to $0$ as $1/L$ , as
expected from the bound discussed below Eq.~\ref{G-average:eqn}.
On the contrary,  when $\ops{H}$ is disordered ($C\to D$ data), 
$[\delta^2_{jj}]_{av}$ converges 
unambiguously to a nonvanishing quantity for $L\to \infty$: time fluctuations survive {\em at all times} 
when $\ops{H}$ is {\em disordered}.
This is in full agreement with the numerical results ~\cite{Gramsch} obtained for the density
after a quench into the localized phase of the Aubry-Andr\'{e} model in one dimension.
%
%
For smaller disorder amplitude $\epsilon$, the situation is similar, except that the 
large-$L$ plateau occurs for larger $L$, due to larger localization lengths.

To better gauge the role of the localized eigenfunctions in making $\delta^2_{j_1j_2}$ finite for $L\to \infty$, 
we have analyzed histograms of the quantities appearing in Eq.~\eqref{eq:delta2}.
A histogram of $|G_{\mu_1\mu_2}|^2$
shows that while the average of $|G_{\mu_1\mu_2}|^2$ scales to zero as $1/L$, see Eq.~\eqref{G-average:eqn}, 
the distribution of 
its values has large tails. 
To analyze these tails, we work with logarithmic distributions, and define
\begin{equation}\label{dist-w:eq}
P_{j_1j_2}^w (x) \equiv  \sum_{\mu_1 \neq \mu_2} \frac{|u_{j_1\mu_1}|^2  |u_{j_2\mu_2}|^2} {\mathcal{N}_{j_1j_2}}
\delta \left( x - \log \left|  G_{\mu_1\mu_2}  \right|^2   \right) \,
\end{equation}
where $\mathcal{N}_{j_1j_2} \equiv \sum_{\mu_1 \neq \mu_2} |u_{j_1\mu_1}|^2 |u_{j_2\mu_2}|^2=1-\sum_{\mu} |u_{j_1\mu}|^2  |u_{j_2\mu}|^2$ is a normalization constant,
$0< \mathcal{N}_{j_1j_2} < 1$, related to the inverse participation ratio \cite{Aoki} when $j_1=j_2$.
%
\begin{figure}[tbh]
   \begin{center}
	\includegraphics[width=0.5\textwidth]{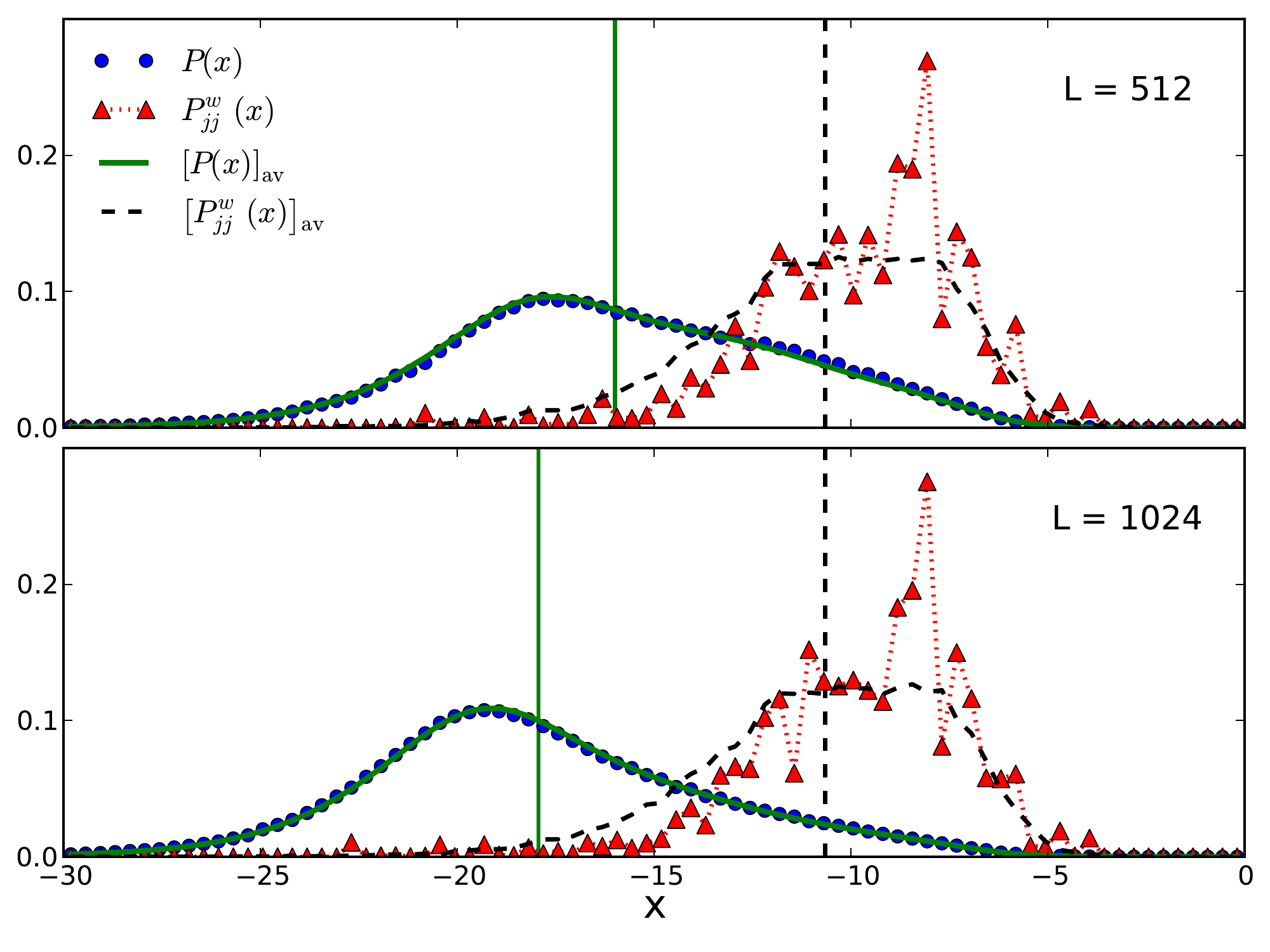}
   \end{center}
\caption{Plot of $P^w_{jj}(x)$, Eq.~\eqref{dist-w:eq} with $j=L/2$ (triangles), and $P(x)$, Eq.~\eqref{dist-unw:eq} (circles),
for  $XX$ chain quenches from a clean $\ops{H}_0$ (with $h_j=0$) to a 
disordered $\ops{H}$ (with $\epsilon=0.3$), for two values of the chain length $L$.
$[P^w_{j_1j_2}(x)]_{av}$ and $[P(x)]_{av} $ (dashed and solid lines) are averages over different realizations of disorder,
with the corresponding vertical lines indicating their mean values $\int \!\ud x \, x [P^w_{j_1j_2}(x)]_{av}$ and
$\int \!\ud x \, x [P(x)]_{av}$.
}
\label{fig:dist}
\end{figure}
%
Figure~\ref{fig:dist} shows $P_{jj}^w (x)$ for 
two different sizes, and compares it with the unweighted distribution 
\begin{equation} \label{dist-unw:eq}
P (x) \equiv \frac{1}{L(L-1)} \sum_{\mu_1 \neq \mu_2}  \delta \left( x - \log \left|  G_{\mu_1\mu_2}  \right|^2  \right) \;.
\end{equation}
We plot both single-instance distributions (solid points) as well as disorder 
average distributions $[\cdots]_{av}$, denoted by lines. 
The unweighted distribution $P (x)$ is smooth, and self-averaging, 
and moves towards smaller mean values when $L$ increases. 
On the contrary, $P_{jj} ^w(x)$ is more structured (single-instance distributions 
depend on the details of the weights  $|u_{j_1\mu_1}|^2  |u_{j_2\mu_2}|^2 $), but 
its mean does not decrease with $L$, 
due to an eigenfunction reweighting of rare events with large values of 
$|G_{\mu_1\mu_2}|^2$. 
Physically, this is quite transparent: similarly to what happens for the inverse participation ratio \cite{Aoki},
localized eigenstates are rather insensitive to the size, while extended states are. 
The fact that the mean $\int \!\ud x \, x P^w_{j_1j_2}(x)$, 
remains finite for $L\to \infty$ for almost all realizations,
is enough to conclude that $\delta^2_{j_1j_2}$ stays finite in the disordered $\ops{H}$ case.
Indeed, Jensen's inequality implies
$\delta^2_{j_1j_2}  = {\mathcal{N}_{j_1j_2}}\langle e^x \rangle_{P_{j_1j_2}^w} \ge {\mathcal{N}_{j_1j_2}} e^{   \langle x \rangle_{P_{j_1j_2}^w} }$
and $\mathcal{N}_{j_1j_2}$ remains finite when $L\to \infty$.

We have shown that microscopic operators have persistent fluctuations 
after a quench to a 
disordered $\ops{H}$.
Such fluctuations could be averaged out if one considers extensive operators involving sums over all sites.
For instance, in a quench to a final disordered Ising chain, while 
the local transverse magnetization $\sigma^z_j(t) = 2 G_{jj}(t) - 1$ has persistent fluctuations, 
the corresponding extensive operator, the total transverse magnetization (per site) 
$\ops{m_z} = L^{-1} \sum_j \ops{\sigma}^z_j$, 
has fluctuations which decrease to $0$ as $L$ is increased, as we have verified. 
Physically, extensive operators effectively perform a {\em self-averaging} of the fluctuations 
$\delta A(t)$, which then vanish in the $L\to \infty$ limit. 
Again, this agrees with the numerical results of Ref.~\cite{Gramsch}.

Let us comment on the issue of thermalization, namely if an ensemble exists which is 
able to describe long-time averages of Ising or $XY$ chains.
One can prove \cite{Ziraldo-2} that for one-body operators (like $G_{j_1j_2}(t)$ and $F_{j_1j_2}(t)$),
independently of the size --- even finite --- and of the quench, time averages coincide with 
the corresponding GGE average: this result can be traced back to the constraints imposed by the GGE.
Thanks to the Wick's expansion, GGE averages reproduce the time averages for any $\ops{A}$ that can 
be expressed as a finite linear combination of powers of $\ops{c}_j$ and $\ops{c}^\dagger_j$, 
{\em as long as the time-fluctuations of $G_{j_1j_2}(t)$ and $F_{j_1j_2}(t)$ vanish}.
On the contrary, when the disorder makes the fluctuations of $G_{j_1j_2}(t)$ and $F_{j_1j_2}(t)$ 
persistent, the time-averages of many-body operators might in general differ from their GGE value. 
We have verified that this is the case, for instance, for $\ops{\sigma}^z_j\ops{\sigma}^z_{j+r}$ 
in quenches from a clean to a disordered $XX$ chain. 
These results are in agreement with the numerical findings of Ref.~\cite{Gramsch}. 

In conclusion, we have shown that the spectral properties and the localization or delocalization 
of the eigenstates of the final Hamiltonian play an important role in characterizing the time fluctuations of observables. 
While the observations we made are based on the analysis of quenches of an integrable
Hamiltonian we expect that the structure of $F_A(\Omega)$ is what ultimately governs the fluctuations: 
integrability should not be crucial in this respect.
%
%
The fact that, in our one-dimensional models, disorder localizes all eigenstates is also likely not crucial.
For fermions hopping in a three-dimensional Anderson model, where there is a mobility edge separating 
localized and delocalized eigenstates, Eq.~\eqref{eq:delta2} is still valid, and we expect 
that localized eigenstates contribute a finite quantity to $\delta^2_{j_1j_2}$ in the thermodynamic limit.
Both these issues, however, clearly call for future investigations.

We thank F. Becca, M. Fabrizio, J. Marino, G. Menegoz, A. Russomanno,
 and P. Smacchia for discussions.
Research was supported by SNSF, through SINERGIA
Project No. CRSII2 136287 1, by the EU-Japan Project
LEMSUPER, and by the EU FP7 under Grant Agreement
No. 280555.

\end{document}